\documentclass{article}

\usepackage{arxiv}

\usepackage[utf8]{inputenc} 
\usepackage[T1]{fontenc}    
\usepackage{hyperref}       
\usepackage{url}            
\usepackage{booktabs}       
\usepackage{amsfonts}       
\usepackage{nicefrac}       
\usepackage{microtype}      
\usepackage{calc}
\usepackage{epsfig}
\usepackage{subcaption}
\usepackage{calc}
\usepackage{amssymb}
\usepackage{amstext}
\usepackage{amsmath}
\usepackage{amsthm}
\usepackage{multicol}
\usepackage{pslatex}
\usepackage{apalike,pifont}
\usepackage{graphicx}
\usepackage{doi}
\usepackage{cite}
\usepackage{amsfonts}
\usepackage{algorithmic}
\usepackage{graphicx}
\usepackage{textcomp}
\usepackage{multirow}
\usepackage{xcolor}
\usepackage{booktabs}
\usepackage{hyperref}
\usepackage{lscape}

\title{Behind the (Digital Crime) Scenes: An MSC Model}

\author{ \href{https://orcid.org/0000-0002-7045-0213}{\includegraphics[scale=0.06]{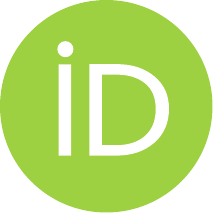}\hspace{1mm}Mario Raciti} \\
	IMT School for Advanced Studies Lucca\\
	Lucca, Italy \\
        Dipartimento di Matematica e Informatica\\
	Università di Catania\\
	Catania, Italy \\
	\texttt{mario.raciti@imtlucca.it} \\
	\And
	\href{https://orcid.org/0000-0002-7615-8643}{\includegraphics[scale=0.06]{orcid.pdf}\hspace{1mm}Giampaolo Bella} \\
	Dipartimento di Matematica e Informatica\\
	Università di Catania\\
	Catania, Italy \\
	\texttt{giamp@dmi.unict.it} \\
}

\renewcommand{\shorttitle}{Behind the (Digital Crime) Scenes: An MSC Model}

\begin{document}
\maketitle

\begin{abstract}
Criminal investigations are inherently complex as they typically involve interactions among various actors like investigators, prosecutors, and defendants. The pervasive integration of technology in daily life adds an extra layer of complexity, especially in crimes that involve a digital element. The establishment of digital forensics as a foundational discipline for extracting digital evidence further exacerbates the complex nature of criminal investigations, leading to the proliferation of multiple scenarios.
Recognising the need to structure standard operating procedures for the handling of digital evidence, the representation of digital forensics as a protocol emerges as a valuable opportunity to identify security and privacy threats. In this paper, we delineate the protocols that compose digital forensics within a criminal case, formalise them as message sequence charts (MSCs), and identify their functional requirements.
\end{abstract}


\section{Introduction}

Criminal investigations have experienced a profound transformation with the integration of digital forensics to address the complexities inherent in modern crime scenarios. Nowadays, digital evidence assumes a central role in legal proceedings, as it introduces new dimensions and challenges that stress the need for a delicate balance between effective investigation and safeguarding the rights of the individuals that are being investigated, namely the defendants. In this context, the rights of defendants are guaranteed by rules and laws ranging at different levels, both national and international. The right to privacy, non-discrimination, information, interpretation, and translation, the right to have a lawyer, the presumption of innocence, and a fair legal process, are merely some examples of the pieces that form the bedrock of a just and equitable legal system~\cite{eurorights, sherlocrights}. Furthermore, the significance of these rights also extends into the digital sphere. An example is given by the European Directive on the protection of personal data for the prevention, investigation, detection, or prosecution of criminal offences~\cite{eurodirective} that emphasises the need for lawful processing, explicit and legitimate purposes, and proportionality in data collection in criminal procedures.

\subsection{Context and Motivation}
\label{subsec:motivation}

Meterko et al.~\cite{meterko} identified cognitive biases, organizational traps, and probability errors that lead to failures in criminal investigations, such as the mistaken prosecution of an innocent person or the failure to prosecute the guilty person. In fact, Rossmo~\cite{rossmo2009criminal} argued that there are three types of criminal investigative failure: (1) ignored crimes, (2) unsolved crimes that should have been cleared, and (3) wrongful convictions. While the second type of failure is the most common, the third is arguably the most damaging. The National Registry of Exonerations has recorded over 3,000 cases of wrongful convictions in the United States as of 2023~\footnote{\url{https://www.ncja.org/crimeandjusticenews/nij-report-the-role-of-forensic-evidence-in-wrongful-convictions}}. Many of these cases involve procedural errors or misconduct by investigators. An example that involved digital forensics is given by the \textit{United States vs. Ganias case}~\footnote{\url{https://www.brennancenter.org/our-work/court-cases/us-v-ganias}}, which involved the seizure and imaging of hard drives during an investigation. The case raised questions about the Fourth Amendment protection from unreasonable search and seizure in the context of digital data, as a confirmation of concerns about data protection principles applied in criminal investigations. Furthermore, the investigators created forensic mirror images of all three of Ganias’s computers but did not purge or delete the files that did not pertain to the case and were therefore non-responsive to the warrant. This led to a debate about the evidentiary chain of custody and the potential creation of a right to deletion, which could complicate criminal prosecutions.
A second example is given by the \textit{United States vs. Comprehensive Drug Testing Inc. case}~\footnote{\url{https://caselaw.findlaw.com/us-9th-circuit/1166919.html}}, where the federal government obtained a subpoena for the drug testing records of ten specific baseball players from a third-party drug testing facility. Yet, instead of limiting their seizure to the records of these ten players, they seized the records of hundreds of other players, who were not identified in the government’s search warrant, nor implicated in any criminal investigation of steroid use. The Ninth Circuit Court of Appeals held that the government was only entitled to information about the ten players it was investigating, and that the rest of the electronic data was obtained illegally.

In brief, a fair and just crime investigation cannot be an arbitrary process, on the contrary it must follow the rules and laws set by a given national or international authority to ensure the defendants' rights. In a cybersecurity fashion, a crime investigation is a protocol where we can identify actors with roles, interactions with exchange of messages, and requirements.
Message Sequence Charts (MSCs) stand as an appealing visual formalism that vividly captures and depicts patterns of interactions within complex systems. 
MSCs are a widely used notation for expressing requirements specifications of multi-agent systems~\cite{anca}. Therefore, MSCs represent a valuable tool to enucleate how digital forensics in crime investigation works. Following this argument, a well-defined and straightforward visual and formal model paves the way to a simplified understanding of the dynamics behind digital crime scenes for the various parties involved, e.g., investigators, prosecutors, defendants. As an additional benefit, a formal and general model will also serve as a baseline reference for threat modelling exercises, hence the general motivation for this paper.

\subsection{Research Questions and Contributions}
\label{sec:contributions}

Following the context and motivations given above, this paper focuses on the formalisation of digital forensics in crime investigation and addresses the following research questions:

\begin{quote}
RQ1 \textit{What are the available documents that explain how digital forensics in crime investigation works?}
\\
RQ2 \textit{Can we extrapolate a general MSC model for digital forensics in crime investigation from that knowledge base?}
\end{quote}

To go about such questions, we search for relevant available documents online, including rules, regulations and best-practices, that focus on digital forensics in crime investigation. Upon establishing a heterogeneous knowledge base from the gathered sources, we elicit fundamental components, shared across different states and authorities, that represent the baseline for the formulation of a generalised MSC model for Digital Forensics in Crime Investigation (DFCI). More specifically, our MSC-model approach comprises a methodical and structured sequence of steps. The main contributions of this paper can be summarised in the steps of our approach:

\begin{enumerate}

\item Identify the Key Actors

\item Identify the Messages

\item Model the Interactions

\item Elicit the Functional Requirements

\end{enumerate}

\section{Related Work}
\label{sec:related-work}

There have been several attempts of modelling digital forensics in literature.
Chun et al.~\cite{chun} studied how to use UML diagrams to model and visualise various aspects of a computer forensics system. Leigland and Krings~\cite{Leigland2004AFO} proposed a formal model for analysing and constructing forensic procedure. Furthermore, Jill et al.~\cite{jill} reviewed the development of digital forensic models, procedures and standards to lay a foundation for the discipline. In addition, Valjarevic and Venter~\cite{valiarevic} emphasised the need for a harmonised digital forensic investigation process model, which can standardise the investigation process and ensure the preservation of evidence integrity. In addition, Montasari~\cite{montasari2016} proposed a model to enable digital forensic practitioners in following a uniform approach when carrying out investigations in the different environments of digital forensics. Subsequently, Montasari et al~\cite{Montasari2019} defined the Standardised Digital Forensic Investigation Process Model (SDFIPM) for conducting digital forensic investigations. Thakar~\cite{Thakar_2021} advanced the Next Generation Digital Forensic Investigation Model (NGDFIM). Moreover, Horseman and Sunde~\cite{horseman} presented the Digital Forensic Workflow Model (DFWM), a novel approach to the structuring and definition of the procedures and tasks involved in the digital forensic investigation process.

These contributions focus on the digital forensics process, yet they lack a specific focus on crime investigation, thus omitting the surrounding context of criminal procedure. To the best of our knowledge, there are no works that focus on a formalisation of digital forensics in crime investigation as an MSC model.

\section{Digital Forensics in Crime Investigation}
\label{sec:comparative-analysis}

As a researcher, we wonder: ``what are the available documents that explain how digital forensics in crime investigation works?''. In the pursuit of a comprehensive understanding of digital forensics in crime investigation, we seek the most relevant available documents online, including rules, regulations and best-practices, that detail how a criminal procedure works, the principal actors and the typical steps, i.e., commencing from the occurrence of a crime and concluding with the issuance of a sentence or verdict, and the inclusion of digital forensics.
In particular, we consider the following sources:

\begin{itemize}
    \item \textit{Criminal Procedure by Wikipedia}~\cite{criminalwiki}. This article offers insights into criminal procedures worldwide, highlighting variations across jurisdictions and delineating fundamental rights in democratic legal systems. It distinguishes between criminal and civil cases, emphasising the initiation of criminal actions by the state.
    \item \textit{Comparative Criminal Procedure by the US Federal Judicial Center}~\cite{comparativecp}. This collection provides an overview of comparative criminal procedure, offering examples from around the world to highlight differences in practices and institutions. It distinguishes between inquisitorial and adversarial legal tradition systems and offers a description of the main steps of a criminal procedure, including initiating a criminal case, investigation, trial. This collection also highlights the main roles that are involved in a criminal procedure, e.g, prosecutor, judge, expert.
    \item \textit{Rights of Defendants (criminal proceedings) by the European Commission}~\cite{eudefrights}. This website provides factsheets per EU member state showing the criminal process and the various steps involved. These factsheets explain defendant's rights and obligations at each stage, from the time of pre-trial investigations, right through to after the trial. In fact, they cover key areas including which authority carries out investigations, how to get legal advice, the roles and rights of the various entities and officials and information on any deadlines that may apply during the process and the assistance available to defendants (including their obligations). The factsheets are divided per EU region.
    \item \textit{Italian Code of Criminal Procedure}~\cite{cpp}. This code contains the rules governing criminal procedure in every court in Italy. The choice of the Italian code as representative stems from a deliberate effort to navigate between the inquisitorial and adversarial legal traditions, as the Italian code could be considered to be somewhere in between the inquisitorial system and the adversarial system.
    \item \textit{How a Criminal Case Works by the UK Crown Prosecution Service}~\cite{ukcps}. This website explains, as suggested by its title, how a criminal case works in England and Wales with a visual format, including key actors and steps.
    \item \textit{Steps in the Federal Criminal Process by the US Department of Justice}~\cite{uscp}. This website illustrates how the criminal process works in the US federal system, explaining the main concepts from the US Federal Rules of Criminal Procedure. As each state has its own court system and set of rules for handling criminal cases, the website also provides some examples of differences between the state and federal criminal processes, including key actors and steps.
    \item \textit{The Budapest Convention (ETS No. 185) by the European Council}~\cite{budapest}. Also known as the Convention on Cybercrime, it is the first comprehensive international treaty that addresses Internet and computer crime by harmonising national laws, improving investigative techniques, and increasing cooperation among nations.
    \item \textit{ISO 27037, ISO 27043}~\cite{iso27037, iso27043}. These standards respectively provide guidelines for specific activities in the handling of digital evidence and for common incident investigation processes involving digital evidence.
    \item \textit{Standard Operating Procedures (SOPs) for the collection, analysis and presentation of electronic evidence by the Council of Europe}~\cite{coe}. This tool on SOPs is intended to provide practical technical and tactical guidance on digital evidence, representing a support to existing rules or, alternatively, a starting point for the development of such procedures for countries or authorities that do not yet have procedures for the collection, analysis and presentation of electronic evidence in place.
    \item \textit{Electronic Evidence Guide Version 3.0 by the Council of Europe}~\cite{eeg}. This guide provides support and guidance to criminal justice professionals on how to identify and handle electronic evidence in such ways that will ensure its authenticity for later admissibility in court. The guide is available upon email request.
    \item \textit{Best Practice Manual for the Forensic Examination of Digital Technology by ENFSI (European Network of Forensic Science Institutes)}~\cite{enfsi}. This document provides technical guidance to aid the design of local standard operating procedures in compliance with local regulatory requirements and international standards.
\end{itemize}

\section{A Generalised MSC Model for DFCI}
\label{sec:model}

From the sources that we discussed in Section~\ref{sec:comparative-analysis}, we can extrapolate and generalise the key actors and main steps, at a high level of generality, involved in digital forensics within criminal procedure, save for the specific terminology of a given state. First, we set the actors. When a crime is reported, a \textit{Prosecutor} investigates on a given \textit{Suspect}, who late would become a \textit{Defendant}. When a crime case necessitates to deal with digital devices to gather potential digital evidence, the figure of the \textit{Digital Forensics Expert} (DF Expert in short) comes into play. Furthermore, legal traditions shape procedure (i.e., investigation, trial, evidence, plea bargaining, appeals) along with the role of the \textit{Judge}. As an example, in the Italian legal system, the actors identified above are reflected as follows: Digital Forensics Expert is ``Consulente Tecnico Informatico'', Prosecutor is ``Pubblico Ministero''), Judge is ``Giudice di Dibattimento'' in later stages), and the Suspect is ``Indagato'', later becoming ``Imputato'' in subsequent phases.

Furthermore, we identify three macro-phases that, together with the key actors, allow us to generalise a DFCI model that is composed of three protocols:

\begin{itemize}
    \item \textbf{Protocol 1: Init} -- It starts from the news of a potential crime and ends before the investigation.
    \item \textbf{Protocol 2: Investigation} -- It embodies the essence of the investigation to find (digital) evidence of the crime.
    \item \textbf{Protocol 3: Trial} -- It represents the final phase of a criminal procedure, terminating with a sentence.
\end{itemize}

While the full detail of each protocol is given in the next sections, the proposed MSC model attains a high level of generality and applicability, as it extrapolates the basic steps of a typical digital forensics process in a criminal procedure from the gathered sources. This makes the model valid and adaptable across different jurisdictions and legal systems, yet under the following hypotheses and assumptions.

\paragraph{The Need for Digital Forensics}
The news of a potential crime can trigger action from the Prosecutor. From this news, we assume that the Prosecutor is able to infer the need for a Digital Forensics Expert to assist the investigative process, given the digital nature of the alleged crime.

\paragraph{The Retention of the DF Expert}
In inquisitorial legal systems, judges appoint impartial experts, often from an official list and commonly employed by national forensic laboratories. Although the prosecution and defence can present their own experts to review findings, these experts may challenge the court expert's assessment. In adversarial systems, both prosecution and defence retain separate experts who testify and prepare reports, with procedural rules governing expert testimony.
In the proposed MSC model, we assume that the DF Expert is retained by the Prosecutor only.

\paragraph{The Phases of Digital Forensics}
There are several definitions for digital forensics. The proposed MSC model inherits the digital forensics phases from the process presented by Palmer~\cite{dfrws} in the First Digital Forensic Workshop, for its widely recognition as a reference and its generality.

\paragraph{The Variety of Trials}
Legal proceedings vary across states, influenced by historical, cultural, and legal traditions. The proposed MSC model simplifies this complexity, assuming a single trial without additional elements like short trials, preliminary hearings, or appeals. The model also omits the inclusion of a jury, concentrating solely on the trial judge as the entity responsible for delivering a final sentence in the case.

\subsection{Protocol 1: Init}

\begin{figure*}[ht]
    \captionsetup{singlelinecheck = false, format= hang, justification=raggedright, font=footnotesize, labelsep=space}
    \centering
    \includegraphics[height=0.37\textheight]{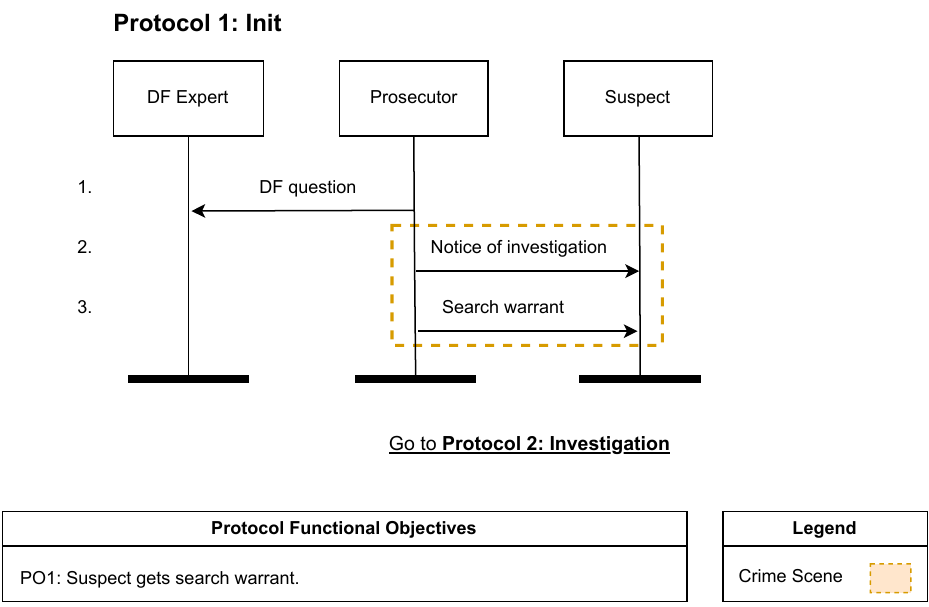}
    \caption{MSC for Protocol 1: Init.}
    \label{fig:protocol1}
\end{figure*} 

The first protocol, illustrated in Figure~\ref{fig:protocol1}, starts from Prosecutor becoming aware of the news of a potential crime, i.e., \textit{notitia criminis}, from a third party. If there is substantial and well-founded reason to proceed the investigation, the Prosecutor (1.) instructs further orders for preliminary investigations. Since the Prosecutor is aware of the need of an expert in the field of digital forensics to support the investigations, they retain a DF Expert for the case (6.). At this point, the focus shifts to the crime scene, where the Prosecutor notifies the Suspect of the investigation (7.) and shows the obtained search warrant to the Suspect (8.).

At this point, the focus shifts to the crime scene, where the Prosecutor notifies the Suspect of the investigation (7.).
\\
The functional objective of the protocol is that the Suspect gets a search warrant.

\subsection{Protocol 2: Investigation}

\begin{figure*}[ht]
    \captionsetup{singlelinecheck = false, format= hang, justification=raggedright, font=footnotesize, labelsep=space}
    \centering
    \includegraphics[height=0.65\textheight]{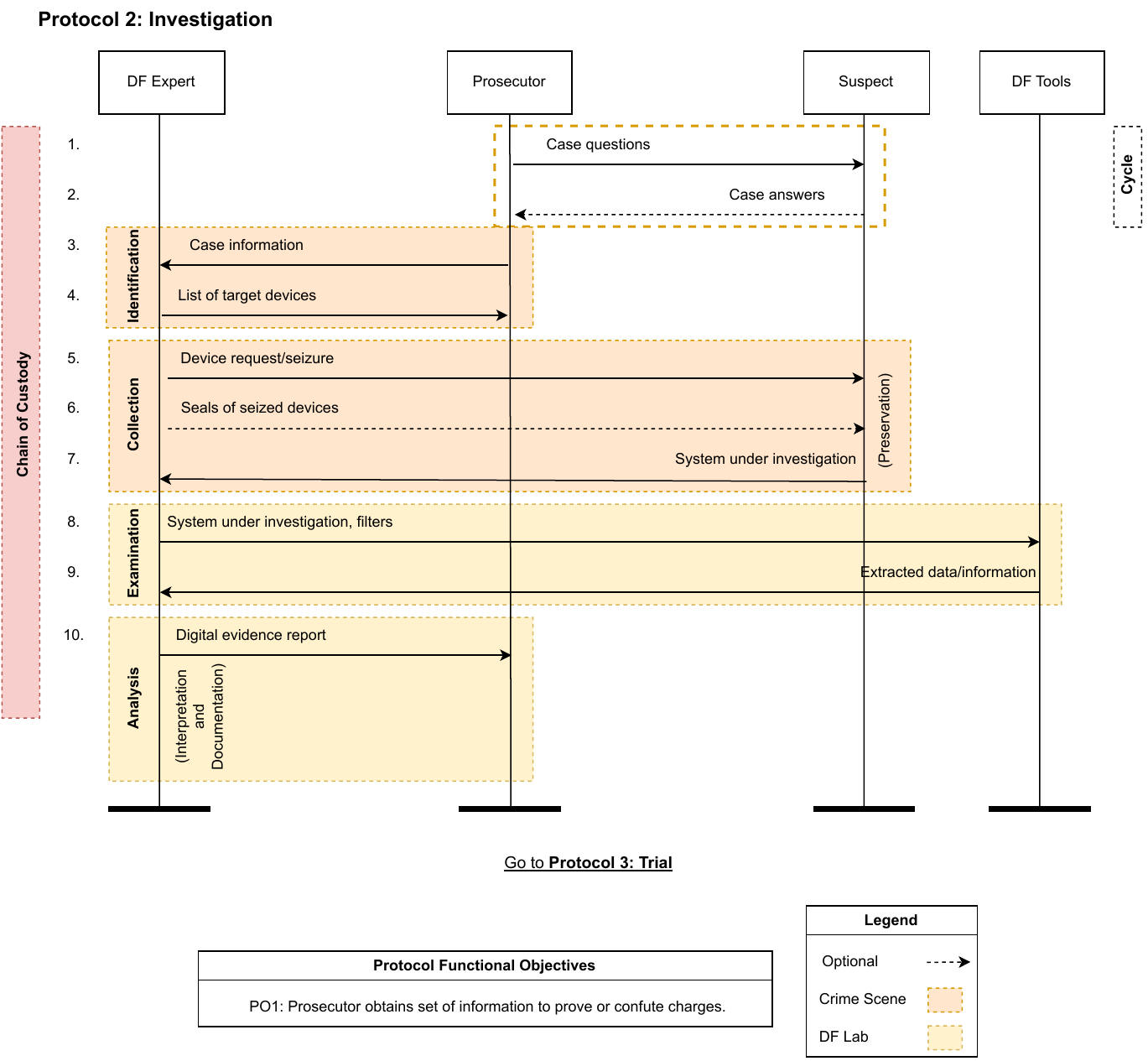}
    \caption{MSC for Protocol 2: Investigation.}
    \label{fig:protocol2}
\end{figure*}

The second protocol, depicted in Figure~\ref{fig:protocol2}, embodies the essence of the investigation to find (digital) evidence of the crime, as it starts from the interrogation of the suspect (1., 2.) and features all the phases of digital forensics -- with the exception of the presentation, which is left for the third protocol. The interrogation phase may involve a cycle.
The DF Expert enters as an active player, starting with the Identification phase, consisting of the Prosecutor giving additional case information to the DF Expert (3.) who, in turn, provides a list of target devices of possible interest for search/seizure (4.). Subsequently, the Collection phase involves the DF Expert requesting/seizing the devices from the Suspect (5.), additionally showing the seals of seized devices to the latter as an optional step (6.). This is suggested by the best-practices, although it is not mandatory. While preserving all gathered devices, the DF Expert receives the so-called system under investigation (7.), i.e., the set of all collected devices, from the Suspect. 

At this point, the focus shifts to the Digital Forensics Laboratory, where the Examination phase is reflected by the DF Expert providing the system under investigation and filters that they deem to be relevant to their DF Tools (8.) which, in turn, return the extracted data/information (9.). DF Tools here represent the set of tools that may be employed by the DF Expert. The last step of the protocol sees the DF Expert forward a digital evidence report, comprising the analysis, interpretation, and documentation of the entire process, to the Prosecutor (10.). Notably, a chain of custody is started and maintained during all the steps of this protocol.
\\
The functional objective of the protocol is that the Police and the Prosecutor obtain a set of information to prove or confute the charges against the Suspect.

\subsection{Protocol 3: Trial}

\begin{figure*}[ht]
    \captionsetup{singlelinecheck = false, format= hang, justification=raggedright, font=footnotesize, labelsep=space}
    \centering
    \includegraphics[height=0.5\textheight]{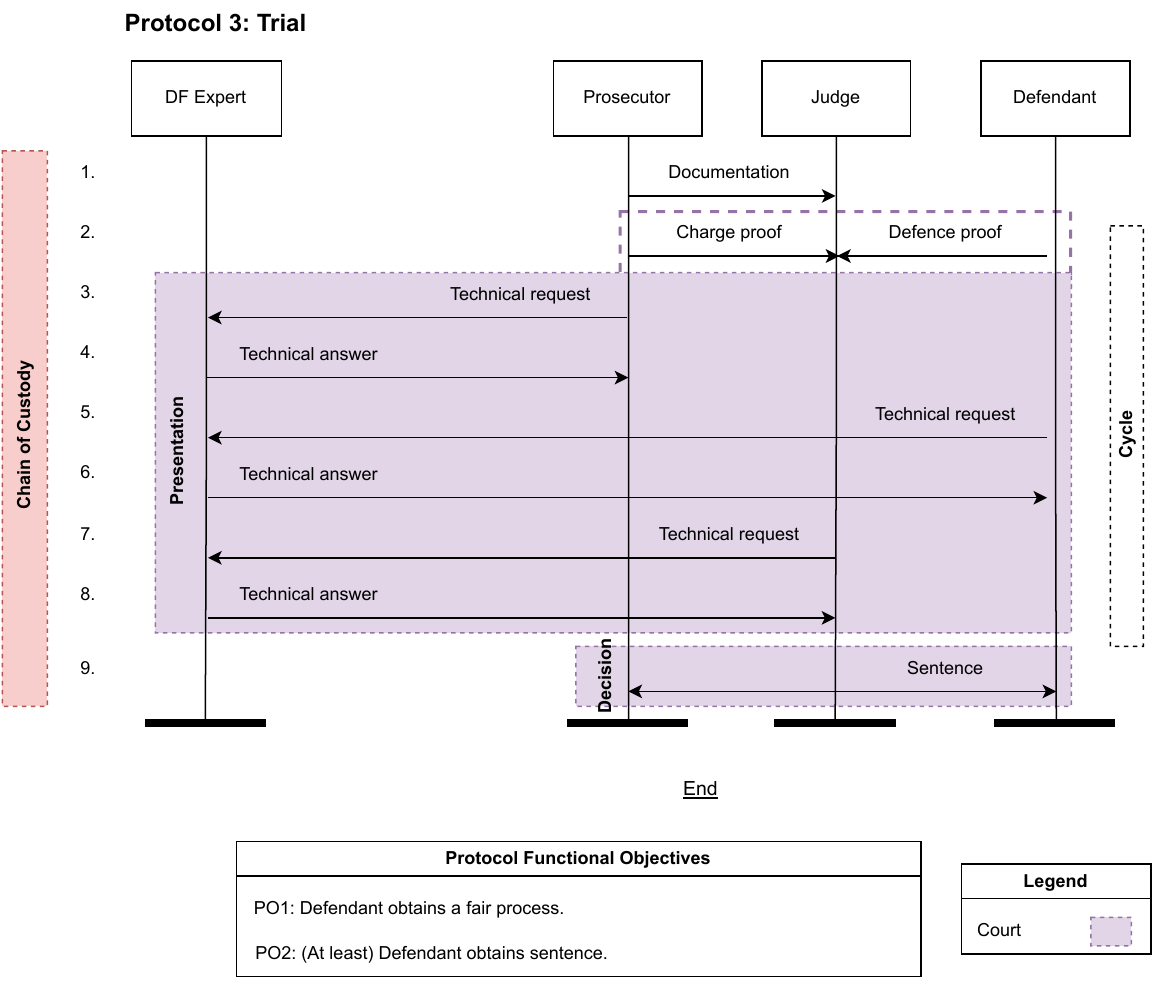}
    \caption{MSC for Protocol 3: Trial.}
    \label{fig:protocol3}
\end{figure*} 

The third protocol, shown in Figure~\ref{fig:protocol2}, concludes with the trial phase. The Prosecutor sends their documentation to the Judge (1.).
At this point, the focus shifts to the court, where both the Prosecutor and the Defendant respectively send the charge proof and defence proof (2.). This starts a possible cycle, where also the Presentation phase of digital forensics is carried out. Each of the main actors in the trial, i.e., Prosecutor, Judge and Defendant, can challenge the DF Expert with a technical request (3., 5., 7.) to receive, in turn, a technical answer (4., 6., 8.) based on the previous (digital) investigation. The protocol ends with a sentence promulgated by the Judge (9.) to the Prosecutor and Defendant, representing the Decision phase. Notably, the chain of custody is maintained till the end of the protocol.
\\
The functional objectives of this protocol are that the Defendant obtains a fair process, and at least the Defendant obtains a sentence.

\section{Discussion}

The proposed MSC model can be used as a reference for cybersecurity research. The three protocols summarise digital forensics in crime investigation under the assumptions discussed in Section~\ref{sec:model}. While we acknowledge potential scepticism regarding the proposed MSC model universality across all states and authorities globally, it reflects an interpretation of digital forensics in crime investigation from the analysis of relevant documents found available online, thus recognising the inherent diversity and dynamism of legal systems worldwide. We cannot claim that the list of sources is exhaustive, yet the different scopes of the documents exhibit a significant coverage for the purpose of general abstraction.

Furthermore, prosecutors, law enforcement, DF experts, and other relevant stakeholders can leverage the proposed MSC model to become aware of potential cybersecurity and privacy issues that may arise, as stressed from the non-functional requirement elicitations.
Moreover, defendants can leverage the MSC model to acquire a knowledge base that helps them to enforce their rights. However, a malicious defendant may exploit this knowledge to develop novel anti-forensics techniques or refine old ones.

\section{Conclusions}

This paper discussed a visual formalisation of digital forensics in crime investigation by relying on the claim that the overall process, from the news of a committed crime to a sentence in a court, is a protocol, as it must adhere to appropriate rules to ensure individuals' fundamental rights.

This paper answered the research questions by gathering available documents that explain how digital forensics in crime investigation works (RQ1) and leveraging it to advance an MSC model (RQ2). In particular, the proposed MSC model details the key steps and phases of a generalised process, including the principal actors and their interactions, hence providing a tool for the elicitation of functional requirements. The high level of generality and applicability of the model is derived from the shared elements in common among the selected document sources.

Our future work looks at refining the proposed MSC model and using it as a reference towards further cybersecurity directions, thus focusing on variations of threat models, elicitation of non-functional requirements, identification of potential attacks that could compromise both the investigative process and/or the defendant's rights, definition of measures to mitigate those attacks. This will hopefully enable different stakeholders, e.g., the scientific community, defendants, law enforcement, digital forensics experts, to look at the digital forensics process in crime investigation from a cybersecurity perspective.

\bibliographystyle{unsrtnat}
\bibliography{references}  






\end{document}